\documentclass[english,aip,jap]{revtex4-1}

\usepackage[T1]{fontenc}
\usepackage[latin1]{inputenc}
\usepackage{amsmath}
\usepackage{graphicx}
\usepackage{amssymb}
\usepackage{dcolumn} 
\usepackage{bm} 
\usepackage{color}

\makeatletter

\usepackage{amsfonts}
\usepackage{epsfig}

\newcommand{\Eq}[1]{Eq.~(\ref{#1})}

\newcommand{\Fig}[1]{Fig.~\ref{#1}}

\usepackage{babel}
\makeatother

\begin{document}

\title{Surface coupling effects on the capacitance of thin insulating films}

\author{Tayeb \surname {Jamali}}
\affiliation{Department of Physics, Shahid Beheshti University, G.C., Evin, Tehran 19839, Iran}

\author{S. \surname {Vasheghani Farahani}}
\affiliation{Department of Physics, Tafresh University, P.O. Box 39518-79611, Tafresh, Iran}

\author{Mona \surname {Jannesar}}
\affiliation{Faculty of Physics, Department of Condensed Matter, University of Kashan, Kashan, Iran }

\author{George \surname {Palasantzas}}
\affiliation{Zernike Institute for Advanced Materials, University of Groningen, Nijenborgh 4, 9747 AG Groningen, The Netherlands}

\author{G.R. \surname {Jafari}}
\email{g\_jafari@sbu.ac.ir}
\affiliation{Department of Physics, Shahid Beheshti University, G.C., Evin, Tehran 19839, Iran}

\date{\today}

\begin{abstract}
A general form for the surface roughness effects on the capacitance of a capacitor is proposed. We state that a capacitor with two uncoupled rough surfaces could be treated as two capacitors in series which have been divided from the mother capacitor by a slit. This is in contrast to the case where the two rough surfaces are coupled. When the rough surfaces are coupled, the type of coupling decides the modification of the capacitance in comparison to the uncoupled case. It is shown that if the coupling between the two surfaces of the capacitor is positive (negative), the capacitance is less (higher) than the case of two uncoupled rough plates. Also, we state that when the correlation length and the roughness exponent are small, the coupling effect is not negligible.

\end{abstract}


\maketitle

\section{Introduction}
\label{sec:intro}
Nowadays widespread improvement in technological devices has offered operational accuracy in addition to miniaturization at submicron length scales. However, size effects due to miniaturization creates issues that are different from what experienced in macroscopic sized devices~\cite{Tesanovic}. The fact of the matter is that by reducing the thickness of a thin film, the physical properties of the system vary. The principle issue here is that as devices tend to smaller dimensions, the coupling between their surfaces which is entangled to the existence of a cross correlation between the surfaces becomes more pronounced. Hence, by considering the fast growth in miniaturizing the devices, more attention should be paid on unignorable coupling effects. When the thickness decreases such that it gets comparable to the mean free path of the electron, the surface roughness comes in to play~\cite{Trivedi, Fishman1}. The surface roughness has been studied in the context of, e.g. electric conductivity~\cite{Fishman1, Fishman2, KBLC,JJVM}, electron localization~\cite{McGurn}, thermal conductivity~\cite{Martin,WW,LC,WNK}, magnetization~\cite{zhao3,STVVB,ARM,CECL,KCCWDYK}, capacitance~\cite{SBP}, etching process~\cite{etch1,etch2}, leakage current~\cite{Zhao}, wave scattering~\cite{Zaman} \& shadowing effects~\cite{Salami2014}, surface growth~\cite{Karda,Heda} \& stochastic processes ~\cite{PhysRevLett.91.226101,PhysRevB.71.155423,Fazeli}, etc.
A feature of surface roughness in the context of capacitance is linked to the fact that charges tend to accumulate on sharper areas. This statement proved adequate for Zhao et al. to show how the capacitance of a parallel plate capacitor increases when one of its plates gets rough~\cite{Zhao}. Hence, it is obvious that the capacitance of a parallel plate capacitor which has two rough surfaces should be further modified. This issue provided the basis of this study; the encountered question here is wether a capacitor with two rough surfaces is equivalent to two capacitors in series each having one rough surface?
It is instructive to state that in case of a non-coupled capacitor, if the capacitor is cut in half, the previous results obtained for a capacitor with one rough surface applies, see~\cite{Zhao}. But when coupling exists, it is expected that the type of coupling between the two rough surfaces affects the physical properties of capacitors. In other words, one can expect to see different results obtained from a correlation or anti-correlation between surfaces. Note that when a surface is undergoing a growth process~\cite{Reis,AFOR,HP}, the upper surface would not forget the previous information of the lower surface. Hence, the existence of correlation between the two surfaces is inevitable. Taking in to account the existing models in application to the configuration of the capacitor under consideration in this work, would lead to the elimination of coupling between the two rough plates of the capacitor. Hence, the coupling effects of the two rough surfaces is the center of attention in this work.

\section{Two bounding coupled rough surfaces; Laplacian solution }
\label{sec:methodology}
\begin{figure}[tb]
\centering
\includegraphics[width=0.7\linewidth]{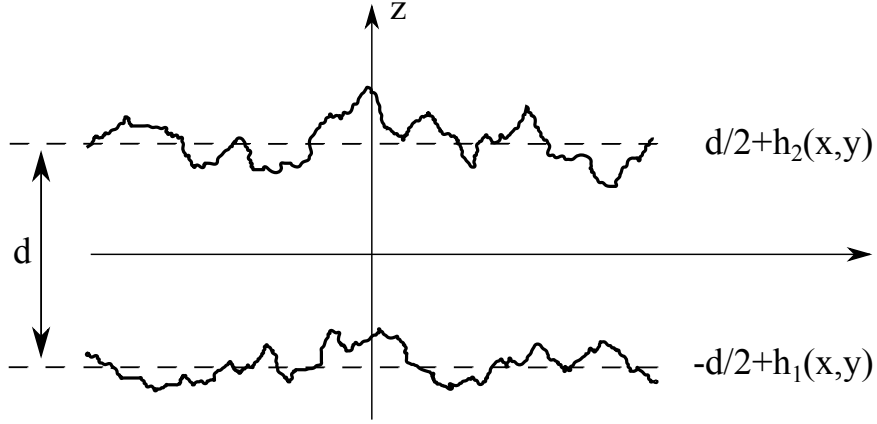}
\caption[Rough capacitor]{Schematic of a parallel-plate capacitor with rough surfaces. The average distance between the plates is $d$, and the surface fluctuations are represented by $h_1(x,y), h_2(x,y)$.}
\label{fig:2}
\end{figure}

Consider a parallel capacitor in which both surfaces are rough with a potential difference of $V$. The average distance between the two rough plates is $d$, where $h_1(x,y),h_2(x,y)$ are the height fluctuations of the lower and upper plates respectively, see~\Fig{fig:2}. The Laplace equation needs to be solved in order to obtain the electrostatic potential $\Phi(x,y,z)$ to provide basis for information about the physical properties of the system
\begin{equation}
\nabla^2\Phi(x,y,z)=0,
\end{equation}
where the boundary conditions for the potential obeys $\Phi(x,y,z=-d/2+h_1(x,y))=0$ and $\Phi(x,y,z=+d/2+h_2(x,y))=V$. It is convenient to expand the boundary conditions using the Taylor expansion
\begin{eqnarray}
\label{eq: Taylor expansion of BCs}
\nonumber
&\Phi &(x,y,z=-d/2+h_1(x,y))=\sum_{k=0}^\infty \frac{1}{k!}h_1^k(x,y)\frac{\partial^k \Phi}{\partial z^k}\bigg|_{z=-d/2}=0, \\
&\Phi &(x,y,z=+d/2+h_2(x,y))=\sum_{k=0}^\infty \frac{1}{k!}h_2^k(x,y)\frac{\partial^k \Phi}{\partial z^k}\bigg|_{z=+d/2}=V.
\end{eqnarray}
Assume that the roughness~\cite{comment1} of the lower and upper surfaces $w_{1,2}$ are small compared to the average distance between the surfaces $d$, therefore it is instructive to utilize the perturbation expansion for the potential as
\begin{equation}
\label{eq: perturbation expantion of potential}
\Phi(x,y,z)=\Phi^{(0)}(x,y,z)+\Phi^{(1)}(x,y,z)+\Phi^{(2)}(x,y,z)+\cdots.
\end{equation}
In this expansion, the $n$th order perturbed potential $\Phi^{(n)}$, has an average as the order of $(w_{1,2}/d)^n$. In order to find the perturbed potentials, we implement the techniques developed by Zhao \emph{et al}~\cite{Zhao}. These perturbed potentials individually satisfy the Laplace equation
\begin{equation}
\label{eq: Laplace eq for petrurbed potentials}
\nabla^2\Phi^{(n)}(x,y,z)=0,
\end{equation}
which is due to the fact that the terms have different orders of magnitude. It is worth stating here that the boundary conditions for the potential $\Phi$ implies boundary conditions on each of the perturbed terms $\Phi^{(n)}$ which can be obtained by substituting Eq.~(\ref{eq: perturbation expantion of potential}) into Eq.~(\ref{eq: Taylor expansion of BCs}) as
\begin{eqnarray}
\label{eq: BCs for perturbed potentials}
\sum_{k=0}^M \frac{h_1^k}{k!}\frac{\partial^k}{\partial z^k}\Phi^{(M-k)} \bigg|_{z=-d/2}=0 \qquad,\qquad \sum_{k=0}^M \frac{h_2^k}{k!}\frac{\partial^k}{\partial z^k}\Phi^{(M-k)} \bigg|_{z=+d/2}=V\delta_{M,0},
\end{eqnarray}
where $M=0,1,2,\cdots$. It should be noted that for each value of $M$, Eq.~(\ref{eq: BCs for perturbed potentials}) would give two recursive relations for the boundary conditions of perturbed potentials. By solving the Laplace equation for each perturbed potential, Eq.~(\ref{eq: Laplace eq for petrurbed potentials}), with the consideration of its specific boundary conditions, Eq.~(\ref{eq: BCs for perturbed potentials}), the full solution for the potential $(\Phi)$ is obtained. Since the surface roughness is assumed to be small compared to the distance between the surfaces, only the first three terms of the perturbation expansion is considered and the rest are neglected.\\
Hence, the zeroth order potential would  be
\begin{equation}
\label{eq:phi0}
\Phi^{(0)}(x,y,z)=\frac{V}{d} \left( z+\frac{d}{2} \right),
\end{equation}
and the first and second order potentials could be obtained making use of the Fourier integral as
\begin{align}
\label{eq:phi1&phi2}
\Phi^{(n)}(x,y,z) &=\int\bigg( A^{(n)}_{+} (\textbf{q}) \frac{\sinh qz}{\sinh \frac{qd}{2}}+A^{(n)}_{-} (\textbf{q}) \frac{\cosh qz}{\cosh \frac{qd}{2}} \bigg)\;\exp(-i\mathbf{q}.\boldsymbol{\rho}) d^2q,
\end{align}
where we have $n=1,2$, $\boldsymbol{\rho}\equiv(x,y)$ and
\begin{align}
\label{eq:A1&A2}
A^{(1)}_{\pm}(\mathbf{q}) &=-\frac{V}{2d} \big( \tilde{h}_2(\mathbf{q})\mp \tilde{h}_1(\mathbf{q})\big), \nonumber \\
A^{(2)}_{\pm}(\mathbf{q}) &=\frac{V}{4d} \int \big[ \tilde{h}_2(\mathbf{q-q'})\mp \tilde{h}_1(\mathbf{q-q'})\big] \big( \tilde{h}_2(\mathbf{q'})- \tilde{h}_1(\mathbf{q'})\big) q' \coth\frac{q' d}{2} d^2 q' \nonumber \\
&\phantom{=}\;  +\frac{V}{4d} \int \big[ \tilde{h}_2(\mathbf{q-q'})\pm \tilde{h}_1(\mathbf{q-q'})\big] \big( \tilde{h}_2(\mathbf{q'})+ \tilde{h}_1(\mathbf{q'})\big) q' \tanh\frac{q' d}{2} d^2 q',
\end{align}
Note that $\tilde{h}(\mathbf{q})=\frac{1}{(2\pi)^2}\int h(\boldsymbol{\rho})e^{i\mathbf{q}.\boldsymbol{\rho}}d^2\rho$ is the Fourier transform of $h(\boldsymbol{\rho})$.

According to \Eq{eq:phi0}, it could readily be noticed that the zeroth order potential is the indication of a smooth parallel-plate capacitor. For the first order potential, as shown in \Eq{eq:phi1&phi2} and \Eq{eq:A1&A2}, the term $(A^{(1)}_{\pm})$ contains information about the height fluctuation $h_i$, where $i=1,2$ ~\cite{comment2}. For the second order potential, as indicated by \Eq{eq:phi1&phi2} and \Eq{eq:A1&A2}, the term $(A^{(2)}_{\pm})$ contains information about the product of the height fluctuations $h_ih_j$, where we have $i,j=1,2$. Strictly speaking, the $n$th order potential consists of terms including the product of the $n$ height fluctuations $h_{i_1}h_{i_2}\cdots h_{i_n}$, where we have $i_1,i_2,\cdots,i_n=1,2$.

\section{Expressions for the electric field and capacitance}
we proceed to find the electric field of our configuration. Since the height fluctuations of the surfaces are considered as a random field obtained by a distribution function, e.g. Gaussian, the potential $\Phi$ would accordingly be a random quantity, where its average at every point inside the capacitor is of our interest. By knowing the potential, the electric field could be obtained. Hence, for the electric field we have
\begin{align}
\label{eq:E}
\mathbf{E}(x,y,z) &=-\nabla\Phi=-\nabla\Phi^{(0)} -\nabla\Phi^{(1)} -\nabla\Phi^{(2)}+\cdots \nonumber \\
&\phantom{=}\;\approx-\frac{V}{d} \hat{\mathbf{e}}_3-\hat{\mathbf{e}}_3\sum_{n=1}^2 \int q  \bigg(  A^{(n)}_{+} (\textbf{q}) \frac{\cosh qz}{\sinh \frac{qd}{2}}+A^{(n)}_{-} (\textbf{q}) \frac{\sinh qz}{\cosh \frac{qd}{2}}\bigg)\exp(-i\mathbf{q}.\boldsymbol{\rho})  d^2q   \nonumber \\
&\phantom{=}\; +i\sum_{n=1}^2 \int \mathbf{q} \bigg(  A^{(n)}_{+} (\textbf{q}) \frac{\sinh qz}{\sinh \frac{qd}{2}}+A^{(n)}_{-} (\textbf{q}) \frac{\cosh qz}{\cosh \frac{qd}{2}}\bigg)\exp(-i\mathbf{q}.\boldsymbol{\rho})  d^2q.
\end{align}
Note that the electric field is a random vector field, and $\hat{\mathbf{e}}_3$ is the unit vector in the $z$ direction. The average of the electric field is also an interest in this work.

In order to obtain the total charge on a rough-surface of a capacitor we recall the fact that the amount of charge accumulated on a typical point of the surface is proportional to the electric field normal to that point. Thus, the total charge accumulated on the lower $(i=1)$ and upper $(i=2)$ rough surfaces is given by $Q_i=\int \mathbf{E}_i.\mathbf{n_i} \; ds$, with surface normal vector $\mathbf{n}_i=(-1)^i\frac{\nabla h_i-\hat{\mathbf{e}}_3}{\sqrt{1+(\nabla h_i)^2}}$. In order to obtain the charge of the plates $(Q_i)$, information about the plates electric field $(\mathbf{E}_i=\mathbf{E}(x,y,{z=(-1)^id/2+h_i}))$ is essential. Note that in order to keep the calculations a bit simple, since the roughness $w_{1,2}$ is smaller compared to $d$, one can write $\mathbf{E}_i\approx\mathbf{E}(x,y,z=(-1)^id/2)$.
Hence, the total charge accumulated in the capacitor is readily obtained. Interestingly the accumulated charge is non-zero, $Q_1+Q_2\neq0$. This is in contrast to Gauss's law for the total charge accumulated on the surfaces of a capacitor; as it speaks for itself it should be zero. Hence, something is odd here. This inconsistency is born from the two different kind of assumed approximations: \textit{(i)} only considering the first three terms of the perturbation expansion in \Eq{eq:E}, and \textit{(ii)} using the approximated version for the plates electric field, $\mathbf{E}_i\approx\mathbf{E}(x,y,z=(-1)^id/2)$. This discrepancy could easily be overcome by taking $\bar{Q}=(Q_2+|Q_1|)/2$,  as magnitude of the charge accumulated on each surface. Hence, the random capacitance is readily obtained
\begin{equation}
\label{eq: capacitance}
C=\bar{Q}/V.
\end{equation}

It is worth noting here that the height fluctuations of the plates behaves in a random manner, so the ensemble average is of interest. Hence, we take the average functional form of the capacitance together with the electric field and potential. The fact of the matter is that in the process of obtaining the ensemble average of $\Phi, \mathbf{E}$ and $C$, the height-height correlation function of the surfaces comes in to play. In the case of a capacitor with only one rough surface it was shown by Zhao \emph{et al}. that for obtaining the ensemble average, the autocorrelation of a rough surface appears only in the second order perturbed term $\Phi^{(2)}$, see ref~\cite{Zhao}. This motivates the study of the cross correlation effects for the case of a capacitor with two coupled rough surfaces. We intend to show that in the averaging process, the cross correlation effects between the two rough surfaces also shows itself in  $\Phi^{(2)}$.\\
There are two basic statistical properties of importance for describing random processes (or random fields). The correlation functions  $R_{ij}(\mathbf{r}_1,\mathbf{r}_2)=\langle h_i(\mathbf{r}_1)h_j(\mathbf{r}_2)\rangle$, and the spectral density functions $S_{ij}(\mathbf{q})=\frac{(2\pi)^3}{A}\langle \tilde{h}_i(\mathbf{q})\tilde{h}_j(-\mathbf{q})\rangle$, where $\langle \cdots \rangle$ denotes the ensemble average over possible roughness configurations. The parameter $A$ denotes the area of the projected plate of the capacitor on the x-y plane, with $i=1,2$~\cite{Bendat2, mandel, Zhao2}.

In the present work, the joint distribution function of the height fluctuation is considered Gaussian with  homogeneous and isotropic surfaces. It is well known that for a homogeneous and isotropic rough surface the correlation functions $R_{ij}(\mathbf{r}_1,\mathbf{r}_2)=R_{ij}(|\mathbf{r}_1-\mathbf{r}_2|)$ and the spectral density functions $S_{ij}(\mathbf{q})=S_{ij}(q)$ are real functions where $q=\vert \mathbf{q}\vert$ ~\cite{Zhao2}. For a homogeneous and isotropic rough surface due to the Wiener-Khintchine theorem~\cite{Bendat2, Zhao2, mandel} the spectral density is the Fourier transform of the correlation function. In addition, one can show
\begin{equation}
\label{eq: generalized Wiener-Khintchine}
\langle\tilde{h}_i(\mathbf{q}_1)\tilde{h}_j(\mathbf{q}_2)\rangle=\frac{(2\pi)^2}{A}\delta(\mathbf{q}_1+\mathbf{q}_2)\langle\tilde{h}_i(\mathbf{q}_1)\tilde{h}_j(-\mathbf{q}_1)\rangle.
\end{equation}

To comply with the aims of this work it is essential to obtain the three main parameters of a capacitor; the average ensemble of the potential, electric field, and capacitance. Making use of Eqs.~(\ref{eq:phi0})-(\ref{eq: generalized Wiener-Khintchine}), and keeping in mind that due to the Wick's theorem for a Gaussian distribution, the ensemble average of the product of any odd number of $\tilde{h}(\mathbf{q})$ is zero, we can obtain
\begin{align}
\label{eq:average phi}
    \langle\Phi\rangle &\approx\Phi^{(0)}+\langle \Phi^{(2)}\rangle \nonumber \\
    &=\frac{V}{d} \left( z+\frac{d}{2} \right)+(2\pi)^2\frac{V\mathcal{P}}{Ad^2}z+(2\pi)^2\frac{V}{2Ad} \int \big( \langle|\tilde{h}_2(\mathbf{q})|^2\rangle-\langle|\tilde{h}_1(\mathbf{q})|^2\rangle\big) q\,\coth qd\,d^2q,
\end{align}
for the potential, and
\begin{align}
\label{eq:average E}
\langle\mathbf{E}\rangle=-\langle\nabla\Phi\rangle\approx-\frac{V}{d} \mathbf{e}_3 -(2\pi)^2\frac{V\mathcal{P}}{Ad^2}\mathbf{e}_3,
\end{align}
for the electric field, and
\begin{align}
\label{eq:average C}
    \langle C\rangle &=\frac{\langle \bar{Q}\rangle}{V}\approx \epsilon_0 \frac{A}{d} \bigg\{ 1+\frac{(2\pi)^2}{Ad} \mathcal{P}+\frac{(2\pi)^2}{2A}\int \bigg(\langle|\tilde{h}_1(\mathbf{q})|^2\rangle+ \langle|\tilde{h}_2(\mathbf{q})|^2\rangle\bigg)q^2\,d^2q \bigg\},
 \end{align}
 for the capacitance. Where
 \begin{align}
 \mathcal{P}&=\int \bigg( \langle|\tilde{h}_1(\mathbf{q})|^2\rangle+\langle|\tilde{h}_2(\mathbf{q})|^2\rangle-\langle\tilde{h}_1(\mathbf{q})\tilde{h}_2(\mathbf{-q})\rangle-\langle\tilde{h}_2(\mathbf{q})\tilde{h}_1(\mathbf{-q})\rangle\bigg) q\,\coth qd \,d^2q.
\end{align}

It could readily be seen in \Eq{eq:average phi} that the autocorrelation and coupling effects show themselves in the term for $\langle\Phi^{(2)}\rangle$. In addition, in the second term of \Eq{eq:average phi}, the contribution of the coupling and autocorrelation are the same in the sense of order. So when the contribution of coupling is as the same order of the autocorrelation, the effects of coupling is not negligible. Note that the last term on the RHS of \Eq{eq:average phi} is just a constant term. \\
In order to obtain the ensemble average of the electric field in \Eq{eq:average E} we use the fact that $\langle\cdots\rangle$ and $\nabla$ can commute with each other. The ensemble average of the electric field has no component on the x-y plane, this is expected for homogeneous and isotropic surfaces. Moreover, $\langle\mathbf{E}\rangle$ is a uniform electric field as in the case for the parallel-plate capacitor. For the capacitance in \Eq{eq:average C} the coupling effect enters in the same order as the  autocorrelations of the surfaces which is similar to that for the potential and electric field.\\

\section{Effects of surface coupling on electrical parameters}
To discuss our results three case studies are carried out; in case of a single rough surface, assuming that the upper surface is rough, the only non-zero term in Eqs.~(\ref{eq:average phi})-(\ref{eq:average C}) is $\langle|\tilde{h}_2(\mathbf{q})|^2\rangle$. This conclusion resembles the results obtained by Zhao \emph{et al}~\cite{Zhao} where the capacitance increases compared to a parallel-plate capacitor due to accumulation on sharp places, see~\Fig{fig:2}. For the case where two uncoupled rough surfaces exist, the two coupling terms $\langle\tilde{h}_1(\mathbf{q})\tilde{h}_2(\mathbf{-q})\rangle$ and $\langle\tilde{h}_2(\mathbf{q})\tilde{h}_1(\mathbf{-q})\rangle$ in Eqs.~(\ref{eq:average phi})-(\ref{eq:average C}) disappear. Thus, the presence of two uncoupled rough surfaces would cause an increase in the $\langle\Phi\rangle, \langle\mathbf{E}\rangle$, and $\langle C\rangle$ compared to a capacitor with only one rough surface. For the case where rough surfaces are coupled, both autocorrelation and coupling terms count. In this case, depending on the functional form of coupling between the two surfaces we experience an increase or decrease in the values of $\langle\Phi\rangle, \langle\mathbf{E}\rangle$, and $\langle C\rangle$ in comparison to two uncoupled rough surfaces.\\
The discussions carried out before this point are general, in a sense that any statement born out in application to the physical quantities of the capacitor up to now disregards the type of roughness of the surfaces. But, since most of the rough surfaces in nature may be considered self-affine, it is best to treat them accordingly, see ~\cite{Palasantzas1} and references therein. An analytic model for the roughness power spectrum which has the proper asymptotic limits that allows calculation of roughness effects is given in ~\cite{Palasantzas1}. This model has the Lorentzian form
\begin{equation}
\label{eq:self affine spec}
\langle |\tilde{h}(\mathbf{q})|^2\rangle=\frac{A}{(2\pi)^3}\frac{w^2\xi^2}{(1+aq^2\xi^2)^{1+\alpha}},
\end{equation}
which proved to be consistent with observed data \cite{Palasantzas1}. The roughness exponent $\alpha$ indicates the degree of roughness irregularity at short length scales $r<\xi$, where $\xi$ is the correlation length ~\cite{barabasi}. The parameter $a$ in \Eq{eq:self affine spec} is introduced in a piece-wise form as
\begin{equation}
a=
\begin{cases}
      \frac{1}{2\alpha}[1-(1+aq_c^2\xi^2)^{-\alpha}] & 0<\alpha<1 \\
      \frac{1}{2}\mathrm{ln}(1+aq_c^2\xi^2) & \alpha=0.
   \end{cases}
\end{equation}
Note that the parameter $q_c=\pi/a_0$ is the upper limit for the frequency in the Fourier space, where $a_0$ is of the atomic order~\cite{Palasantzas1}. It could be noticed from \Eq{eq:self affine spec} that for a self-affine surface, the three parameters $\alpha,\xi$, and $w$, specify all information about the surface. However these parameters could be different for various surfaces.\\
In order to show the effect of coupling between two rough plates we suppose the surfaces self-affine. Consider two uncoupled self-affine rough surfaces characterised by $(\alpha_1,\xi_1,w_1)$ and  $(\alpha_2,\xi_2,w_2)$. The spectral density of each surface is obtained by \Eq{eq:self affine spec} where the coupling terms disappear as discussed earlier. If we take the roughness of the two surfaces as $w_1=w_2=w$, and the roughness exponent of the two surfaces as $\alpha_1=\alpha_2=\alpha$ where $w$ and $\alpha$ are constant, the only variable that remains is the correlation lengths, $\xi_{1,2}$. In this stage it is convenient to perform a change of variable and take $\Delta=w/d, L_{1,2}=\xi_{1,2}/d$ and $q'=qd$ ~\cite{Zhao}. By substituting the spectral density functions of each self-affine surface in \Eq{eq:average E}, the electric field is obtained
\begin{equation}
\label{eq:<E> for two ind}
\frac{\langle\mathbf{E}\rangle_\mathcal{I}}{E_0}=1+\Delta^2\sum_{i=1}^2 L_i^2\int_0^{q'_c}\frac{q'^2 \coth q'}{(1+q'^2L_i^2/2\alpha)^{1+\alpha}}\; dq',
\end{equation}
where the index $\mathcal{I}$  stands for two independent or uncoupled surfaces, and $E_0=V/d$ is the electric field of a parallel-plate capacitor. Similarly the capacitance is
\begin{equation}
\label{eq:<C> for two ind}
\frac{\langle C\rangle_\mathcal{I}}{C_0}=\frac{\langle\mathbf{E}\rangle_\mathcal{I}}{E_0}+\frac{\Delta^2}{2}\sum_{i=1}^2 L_i^2\int_0^{q'_c}\frac{q'^3}{(1+q'^2L_i^2/2\alpha)^{1+\alpha}}\; dq',
\end{equation}
\begin{figure}[tb]
\centering
\includegraphics[width=0.7\linewidth]{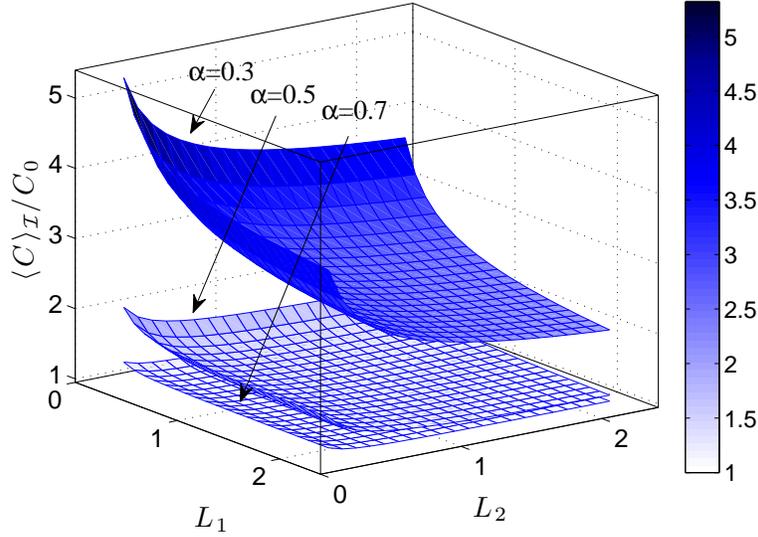}
\caption[Rough capacitor]{Three Surface plots showing the ratios of the capacitances $\langle C\rangle_\mathcal{I}/C^{(0)}$ for the case of two uncoupled rough surfaces in terms of the normalised correlation lengths $L_{1,2}(=\xi_{1,2}/d)$ . From top to bottom the roughness exponent has been taken equal to $0.3, 0.5, 0.7$, with $\Delta=w/d=0.01$.}
\label{fig:3}
\end{figure}
where $C_0=\epsilon_0 A/d$ is the parallel-plate capacitance. Figure~\ref{fig:3} demonstrates the dependence of $\langle\mathbf{C}\rangle_\mathcal{I}/C_0$ on the normalized correlation lengths $L_{1,2}$. Figure~\ref{fig:3} is plotted for three different values of $\alpha$ with $\Delta=0.01$. It could be deduced from Fig.~\ref{fig:3} that for a fixed normalized correlation length $L_{1,2}$, as the roughness exponent $\alpha$ decreases, the ratio $\langle\mathbf{C}\rangle_\mathcal{I}/C_0$ increases. In addition, for a fixed roughness exponent $\alpha$, as $L_1$ or $L_2$ increases, the ratio $\langle\mathbf{C}\rangle_\mathcal{I}/C_0$ decreases to unity.\\
In the case of two coupled rough surfaces, the discrepancy lies in the fact that there is a non-zero cross-spectral density. In general, cross-spectral density $\langle\tilde{h}_1(\mathbf{q})\tilde{h}_2(\mathbf{-q})\rangle$ is equal to
\begin{equation}
\gamma_{12}(\mathbf{q})\sqrt{\langle|\tilde{h}_1(\mathbf{q})|^2\rangle\langle|\tilde{h}_2(\mathbf{q})|^2\rangle},
\end{equation}
where $\gamma_{12}(\mathbf{q})$ called the coherence function is a complex function located in the unit circle of the complex plane~\cite{Bendat1,palasantzas2}. For two homogeneous and isotropic surfaces, the coherence function ($\gamma_{12}(\mathbf{q})=\gamma_{21}(\mathbf{q})=\gamma(q)$) is real belonging to the domain $[-1,1]$. Here, we suppose the simple case $\gamma(q)=-1$ which indicates the negative cross-correlation between surfaces. Hence implementing the same change of variables for the roughness and correlation lengths, we notice that $\left(\langle\mathbf{E}\rangle_\mathcal{C}-\langle\mathbf{E}\rangle_\mathcal{I}\right)/E_0=\left(\langle\mathbf{C}\rangle_\mathcal{C}-\langle\mathbf{C}\rangle_\mathcal{I}\right)/C_0$ equals
\begin{align}
\label{eq:difference}
L_1 L_2 \int_0^{q'_c}\frac{2\Delta^2\;q'^2 \coth q'\;dq'}{[(1+q'^2 L_1^2/{2\alpha})(1+q'^2 L_2^2/{2\alpha})]^{(1+\alpha)/2}},
\end{align}
\begin{figure}[tb]
\centering
\includegraphics[width=0.7\linewidth]{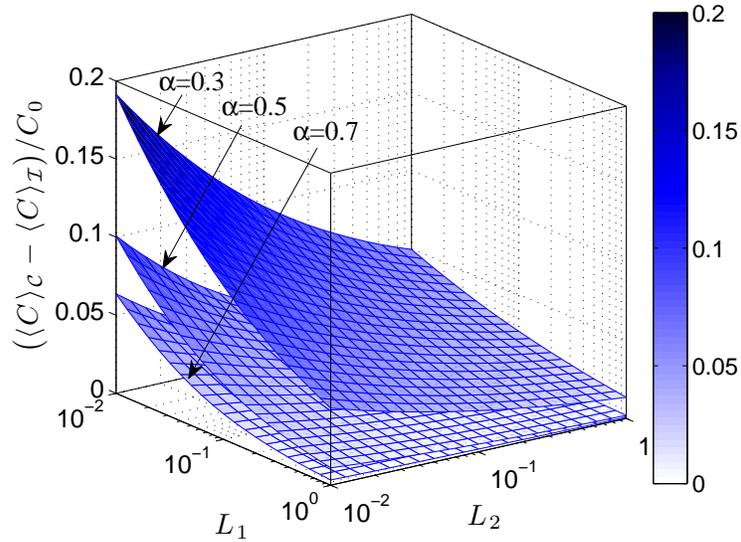}
\caption[Rough capacitor]{Three surface plots illustrating deviation of the coupled capacitance from the uncoupled capacitance $\big(\langle C\rangle_\mathcal{C}-\langle C\rangle_\mathcal{I}\big)/C^{(0)}$ in terms of the normalised correlation lengths $L_{1,2}(=\xi_{1,2}/d)$. From top to bottom the values of the roughness exponent is taken equal to $0.3, 0.5, 0.7$, and we have taken $\Delta=w/d=0.01$. The same goes for the deviation of the corresponding electric fields $\big(\langle E\rangle_\mathcal{C}-\langle E\rangle_\mathcal{I}\big)/E^{(0)}$.}
\label{fig:4}
\end{figure}
where the index $\mathcal{C}$ indicates two \textit{coupled} surfaces. Figure~\ref{fig:4} shows the coupling effect for both the electric field and capacitance~\Eq{eq:difference} as a function of the normalized correlation lengths $L_{1,2}$ for various roughness exponents $\alpha$ at fixed $\Delta=0.01$. Also, it could be deduced that for a fixed normalized correlation length $L_{1,2}$, as the roughness exponent ($\alpha$) decreases, the capacitance of the coupled capacitor increases significantly in comparison to a smooth parallel-plate capacitor.
Moreover, for a fixed $\alpha$, as $L_1$ or $L_2$ increases, a decrease is seen in $\langle\mathbf{E}\rangle_\mathcal{C}$ and $\langle C\rangle_\mathcal{C}$ of the coupled capacitor. In other words, as the normalized correlation lengths prolongs, the system would tend to the case of an uncoupled rough capacitor. It could readily be noticed from~\Fig{fig:4} that the coupling effect is more efficient when the normalised correlation length and the roughness exponent are small.

\section{Conclusions}
\label{sec:conclusions}
In this work the Laplace equation was solved for a parallel-plate capacitor with two rough surfaces. Assuming that the roughness of both surfaces are small compared to the average distance between the surfaces, the perturbation of the electric potential was substituted in the Laplace equation. Since the fluctuations of the surface height was considered very small compared to the width of the capacitor, only up to the second order terms were kept. \\
Solutions to the Laplace equation giving the electric potential of a capacitor with two rough surfaces (either coupled or uncoupled) leads to the conclusion that the electric field and consequently the capacitance increases compared to the case of the smooth parallel-plate capacitor, see also~\cite{Zhao}. This could be explained by the fact that electric charges accumulate on sharp places of a substrate where in the particular case studied here is the summit and foothill of the height fluctuations on the surface. The comparison of a capacitor with two uncoupled rough surfaces with another capacitor with two coupled rough surfaces showed that the increase and decrease of the capacitance for the two cases depend on the sign of their cross correlation. If the two rough surfaces are correlated/anti correlated, the capacitance is decreased/increased (\Eq{eq:average C}) in comparison to the capacitor with two uncoupled rough plates. Note that when the normalised correlation length and the roughness exponent are small, the coupling effect is not negligible, see \Fig{fig:4} \\
The model considered in this work was based on the fact that in a rough surface capacitor, the coupling between the surfaces causes a deviation in its capacitance in comparison to what obtained by entering a slit in between its plates. In other words, the slit creates two capacitors in series (each with one rough surface) where their equivalent capacitance varies from the initial capacitance that had two coupled rough surfaces. The reason for this most interesting deviation is linked to the coupling between the two rough surfaces. To be more precise, the deviation would be most pronounced when the coupling between the rough surfaces is strong. The term strong comes from the fact that the cross correlation of the two rough surfaces is as of the same order of the height-height auto correlation of each rough surface. The contribution of a strong correlation towards the physical parameters of a capacitor could rise up to twenty percent or even more. Hence, considering a capacitor with two coupled rough surfaces as two capacitors in series where each has one rough surface may not be the best assumption. Hence, we understand now that the coupling effects should be taken more seriously into account as devices tend
to miniaturize down to submicron ranges.

\bibliography{rough_surface}

\begin{thebibliography}{39}
\bibitem [{com({\natexlab{a}})}]{comment1}%
  \BibitemOpen
  \href@noop {} {}\bibinfo {note} {Roughness of a surface with the height
  fluctuation $h$ is defined as $w=\sqrt{\langle h^2\rangle-\langle
  h\rangle^2}$}\BibitemShut {NoStop}%
\bibitem [{com({\natexlab{b}})}]{comment2}%
  \BibitemOpen
  \href@noop {} {}\bibinfo {note} {In this letter, the subscript ``1'' or ``2''
  for any quantity represents that quantity for lower or upper
  surface.}\BibitemShut {Stop}%
\end{thebibliography}

\end{document}